\documentstyle[prl,aps,multicol,psfig]{revtex}   % XXX this is new

\newcommand{\ie}{i.e.}
\newcommand{\ea}{\textit{et al.}}

\begin{document}
\draft

\title{Scaling of Crack Surfaces and Implications on Fracture Mechanics}

\author{St\'{e}phane Morel$,^1$ Jean Schmittbuhl$,^2$ Elisabeth Bouchaud$,^3$ 
        and G\'{e}rard Valentin${\,}^1$}
\address{$^1$ Lab. de Rh\'{e}ologie du Bois de Bordeaux, 
        UMR 5103, Domaine de l'Hermitage,
        B.P.10, 33610 Cestas Gazinet, France\\
        $^2$ Lab. de G\'{e}ologie, UMR 8538, Ecole Normale Sup\'{e}rieure,
        24 rue Lhomond, 75231 Paris Cedex 05, France\\
        $^3$ C.E.A.- Saclay (DSM/DRECAM/SPCSI), 
        91130 Gif-Sur-Yvette Cedex, France}
        
\maketitle

\begin{abstract}
  The scaling laws describing the roughness development of crack
  surfaces are incorporated into the Griffith criterion.  We show
  that, in the case of a Family-Vicsek scaling, the energy balance
  leads to a purely elastic brittle behavior.  On the contrary, it
  appears that an anomalous scaling reflects a \textit{R}-curve
  behavior associated to a size effect of the critical resistance to
  crack growth in agreement with the fracture process of heterogeneous
  brittle materials exhibiting a microcracking damage.
\end{abstract}
\begin{multicols}{2}      
%\narrowtext
%
%%%%%%%%%%%%%%%%%%%%%%%%%%%%%%%%%%%%%%%%%%%%%%%%%%%%%%%%%%%%%%%%%%%%%%%%%%%%%
%
\newpage
Fractography has always been a useful tool for the understanding of
fracture mechanisms in heterogeneous materials.  It is well
established that the fracture properties of brittle heterogeneous
materials such as rock, concrete, tough ceramics, wood and various
composites strongly depend on the microstructure and on the damage
process of the material.  Both these parameters are also at the source
of the roughness of crack surfaces.  Indeed, microstructural
heterogeneities acting on the crack front as obstacles, and the
existence of a large fracture process zone with cracking damage where
elastic interactions occur between cracks, may have a strong influence
on the local deviations of the main crack \cite{Hor87,Kac94,Law93}.
Thus, the roughness of fracture surfaces can be considered as an
inheritance of the heterogeneous character and of the damage process
of the material.  Hence, it is naturally tempting to correlate the
fracture surface morphology to macroscopic mechanical properties such
as fracture energy or toughness.
 
The studies of quantitative fractography are nowadays a very active
field of research.  Many experiments \cite{Dag97,Lop98,Mor98,Bou97} on
materials as different as ductile aluminium alloys to brittle
materials like rock or wood have shown that the topography of fracture
surfaces is self-affine \cite{Bar95}.  However, in most cases and
in spite of strong differences in the materials, a roughness index
${\zeta}_{loc}=0.8-0.9$ (called local roughness exponent in what
follows) has been reported in all cases.  This robustness of the
results seems to support the idea suggested by Bouchaud 
\ea~\cite{Bou90} that the local roughness exponent might have a 
universal value, \ie, independent of the fracture mode and of the 
material.

% XXX this is new
Nevertheless, if the local roughness exponent is universal, the range 
of lengthscales within which this scaling domain is observed strongly 
depends on the material microstructure.  Thus, recent studies focussed on 
the complete description (3D) of the crack morphology of granite 
\cite{Lop98} and wood \cite{Mor98} have shown that the scaling 
laws governing the crack developments in longitudinal and transverse
directions are different and material dependent.  
%  
%On the other hand, it has been suggested by Bouchaud \ea~
%\cite{Bou93b} that analytical models of interfacial growth
%\cite{Bar95,Vic92} or front lines propagating through a field of
%randomly distributed obstacles \cite{Nar93,Ert94} might be
%relevant to understand the morphology of fracture surfaces
%\cite{Schm95b,Ram97}.
% 
Let us consider the development of a fracture surface from a border
where a straight notch of length $L$ imposes a zero roughness. The
mean plane of the crack surface is defined as $(x,y)$ where the $x$
axis is perpendicular to the direction of crack propagation and the
$y$ axis is parallel to this propagation direction. It has been found
that the height fluctuations $\Delta h$ of fracture surfaces of two 
heterogeneous brittle materials (granite \cite{Lop98} and wood 
\cite{Mor98})
%
% XXX this is new
%
estimated over a window of size $l$ along the $x$ axis and at a 
distance $y$ from the initial notch 
exhibited \textit{intrinsically anomalous} scaling properties obtained
in some models of nonequilibrium kinetic roughening \cite{Lop97} :
\begin{eqnarray}
        \Delta h(l,y)\simeq A
        \left\{ \begin{array}{ll}
        l^{{\zeta}_{loc}} \: \xi (y)^{\zeta -{\zeta}_{loc}}  
        & \mbox{if \quad $l \ll \xi (y)$} \\
        \xi (y)^{\zeta} & \mbox{if \quad $l \gg \xi (y)$}
\label{Anomalous}
                \end{array}
\right.
\end{eqnarray}
where $\xi(y)=By^{1/z}$ depends on the distance to the initial notch
$y$ and corresponds to the crossover length along the $x$ axis below
which the surface is self-affine with the local roughness exponent
${\zeta}_{loc}$.  Along the $y$ axis, the roughness develops according 
to two different regimes : for large lengthscales, \ie~$l \gg \xi (y)$, 
the roughness grows as $\Delta h(l,y) \sim y^{\zeta/z}$ where $\zeta$ 
is called the global roughness exponent while for small lengthscales, 
\ie~$l \ll \xi (y)$, the roughness growth is characterized by the 
exponent $(\zeta-{\zeta}_{loc})/z$.  
%
% XXX this is new
%
The global roughness exponent $\zeta$, the exponent $z$ (called dynamic 
exponent) and the prefactors $A$ and $B$ are material dependent.  Thus, 
despite the universal distribution of the height fluctuations in the 
transverse direction of crack propagation, the roughening in longitudinal 
direction is material dependent and has an influence on the roughness 
magnitude in the transverse direction.  
All lengths are dimensionless. Real lengths are divided by a unity length
$l^*$.

One main consequence is that, when the global saturation occurs,
\ie~far from the notch (for $y \gg y_{sat}$, where $y_{sat}=(L/B)^z$),
the magnitude of the roughness is not just a function of the window
size but also of the system size $L$ :
%
% XXX this is new
%
$\Delta h(l,y \gg y_{sat}) \simeq A \: l^{{\zeta}_{loc}} \: 
L^{\zeta -{\zeta}_{loc}}$.  
This unconventional dependence of the stationnary local fluctuations
on the system size is distinctly different from what happens in the
Family-Vicsek scaling \cite{Fam91} which is defined as :
\begin{eqnarray}
  \Delta h(l,y) \simeq A
        \left\{ \begin{array}{ll}
        l^{{\zeta}_{loc}}   & \mbox{if \quad $l \ll \xi (y)$} \\
        \xi (y)^{{\zeta}_{loc}} & \mbox{if \quad $l \gg \xi (y)$}
\label{Family}
                \end{array}
\right.
\end{eqnarray}
The Family-Vicsek scaling can be seen as a particular case of anomalous
scaling where: $\zeta={\zeta}_{loc}$.

In this study, we propose to discuss the link between the roughening
of crack surfaces and the fracture process.  On the basis of the
Griffith criterion, we show that a fracture surface exhibiting an
anomalous scaling reflects fundamental mechanical behavior of
quasi-brittle materials: 
%
% XXX this is new
%
\textit{R}-curve behavior (\ie~an evolution of the resistance to crack 
growth as a function of the crack length increment \cite{Law93}) and size 
effect of the elastic energy release rate.  
Fig.~1 shows an illustration of a \textit{R}-curve behavior observed 
for a wood crack propagation (mode I) using a TDCB technique 
%
% XXX this is new
%
\cite{Mor98,Mo-Sc-Bo-Va00}.  
The energy release rate $G$ (proportional to the square of the 
toughness) evolves very significantly at the onset of crack 
propagation and becomes independent of the crack length after a 
characteristic propagation distance. In Fig. 2, an example of size 
effect on this critical energy release rate at saturation is presented 
\cite{Mo-Sc-Bo-Va00}.  
%
%%%%%%%%%%%%%%%%%%%%%%%%%%%%%%%%%%%%%%%%%%%%%%%%%%%%%%%%%%%%%%%%%%%%%%%%%%%%%
%

Several connections between the fractal dimension of the crack
surfaces and the fracture energy or toughness have been proposed
\cite{Man84,Mec89,Mos93,Bou94,Car94,Bal96,Bor97,Baz97}.  Nevertheless, 
most of the proposed analytical models
\cite{Mos93,Bou94,Car94,Bal96,Bor97,Baz97} show weak connections with
the fracture behavior of brittle heterogeneous materials and
especially with the phenomenological \textit{R}-curve behavior and
size effect.  
%
% XXX this is new
%
However, these models have been built on the basis of a 2D self-affine 
crack surface disregarding the anisotropy of the scaling laws governing 
the crack developments in longitudinal and transverse directions.  
In this case, either a \textit{R}-curve is obtained without any size 
effect or a size effect without \textit{R}-curve.   
Let us consider a semi-infinite linear elastic specimen of thickness
$L$ containing an initial crack at position $\Delta a$ submitted to an
uniaxial stable and slow tension (mode I).  In the classical Griffith
approach, the fracture criterion is estimated by balancing the elastic
energy released at the macroscale during an infinitesimal crack
propagation and the energy required to create the corresponding free
surfaces at the microscale.  According to linear elastic fracture
mechanics (LEFM), the rate of energy dissipation in the structure as 
a whole must be defined with respect to the projected crack surface 
$A_p$ (at macroscale, crack roughness can be ignored).  
On the contrary, the estimate of the energy required for crack 
propagation at the microscale needs to take into account the real crack 
surface $A_r$.  Introducing the elastic energy release rate $G$ at the 
macroscale, the fracture criterion becomes for an infinitesimal crack 
advance :
\begin{equation} 
    G \ \delta A_p = 2\gamma\ \delta A_r
\label{Crit}
\end{equation}
where $\delta A_r$ is the real area increment and $\delta A_p$ its
projection on the fracture mean plane.  The term $\gamma$ is the
so-called specific surface energy that characterizes the resistance of
material to cracking.  The real area can be obtained from the surface 
description during the crack growth $\delta a$ from position 
$\Delta a$ of the crack front : 
%
% XXX this is new
%
$\delta A_r\simeq\int_{\Delta a}^{\Delta a+\delta a} \psi (y) \: dy$, 
where $\psi(y)$ is the length of a profile at position $y$ of the
crack roughness.  For a specimen of width $L$, a first order 
approximation leads to the following expansion of the fracture 
criterion: $ G = 2\gamma \psi(y) / L$.  
According to (\ref{Anomalous}), the roughness $\Delta h(l,y)$ can be
seen as a description of the fracture profile at position $y$.  The
length $\psi$ can be estimated by covering the profile
path with segments of length $\delta$ the horizontal projection on
the $x$ axis is $l_o$ and the vertical one on the $z$ axis is on 
average [see Eq.(\ref{Anomalous})] : 
$A \: \xi (y)^{\zeta-{\zeta}_{loc}} \: {l_o}^{{\zeta}_{loc}}$.  
The length $l_o$ is chosen as the lower
cutoff of the fractal range ( \ie~the characteristic
size of the smaller micro-structural element relevant for the fracture
process). Moreover, the crack profile is a univalued function (observed
fracture surfaces do not show overhangs) and hence the number of
segments $\delta$ needed to cover the curve is given by the
ratio $(L/l_o)$.  As a consequence the real length $\psi
=(L/l_o)\delta$ of the crack profile appears dependent on the distance
$y$ to the initial notch and can be expressed as :
\begin{eqnarray}
        \psi (y) \simeq L
        \left\{ \begin{array}{ll}
        \left[ 1+\left( \frac{A(B y^{1/z})^{\zeta-{\zeta}_{loc}}}
        {{l_o}^{1-{\zeta}_{loc}}} \right)^2 \right]^{1/2}
        & \mbox{if $y \ll y_{sat}$} \\
        \left[ 1+\left( \frac{A L^{\zeta-{\zeta}_{loc}}}
        {{l_o}^{1-{\zeta}_{loc}}} \right)^2 \right]^{1/2}
        & \mbox{if $y \gg y_{sat}$}
\label{Length}
                \end{array}
\right.
\end{eqnarray}
In Eq.(\ref{Length}), the terms between brackets correspond to the
tangent of the typical angle between the segment $\delta$ and the $x$
axis. Thus, as expected intuitively, the length of a self-affine curve
is larger than its projected length although it is proportionnal to
it.  On the other hand, according to (\ref{Length}) when the global
saturation state of the roughness is reached (\ie~for $y \gg y_{sat}$), 
the length of the crack profile saturates and 
becomes a nonlinear function of the specimen width $L$.  Note that 
the length of the self-affine crack profile [Eq.(\ref{Length})] 
is a finite quantity in contrast with the expressions proposed by
Carpinteri \cite{Car94}, Borodich \cite{Bor97} and Ba\v{z}ant
\cite{Baz97}.  In these models, the actual length of a fractal curve 
is considered as infinite and this leads the use of an unconventional 
definition of fracture energy (\textit{fractal fracture energy}).  
As we will show in the following, the introduction of a lower cutoff 
for the surface description allows to obtain a fracture energy in 
agreement with the classical dimensions of LEFM.

In the zone where the roughness grows (\ie~for 
$\Delta a \ll {\Delta a}_{sat}$ with ${\Delta a}_{sat}=y_{sat}$), 
the fracture equilibrium (\ref{Crit}) leads to an energy 
release rate function of $\Delta a$ : 
\begin{equation}
        G_R(\Delta a \ll \Delta a_{sat}) \simeq  2 \gamma 
        \sqrt{1+\left( \frac{A B^{\zeta-{\zeta}_{loc}}}
        {{l_o}^{1-{\zeta}_{loc}}} \right)^2  
        {\Delta a}^{2(\zeta-{\zeta}_{loc})/z}}
\label{Gr}   
\end{equation}
In Eq.(\ref{Gr}), the subscript $R$ emphasizes that the behavior of
the resistance to fracture growth is similar to a \textit{R}-curve
\cite{Law93}. When the crack increment is large, \ie~for
$\Delta a \gg {\Delta a}_{sat}$ which corresponds to the saturation
state of the roughness, the resistance to fracture growth becomes :
\begin{equation}
        G_{RC}(\Delta a \gg {\Delta a}_{sat}) \simeq 2 \gamma 
        \sqrt{1+\left( \frac{A}
        {{l_o}^{1-{\zeta}_{loc}}} \right)^2  
        L^{2(\zeta-{\zeta}_{loc})}} 
\label{Grc}  
\end{equation}
In the zone where the roughness saturates, the length of the crack 
profile remains constant which induces a resistance to crack
growth independent of the crack length increment $\Delta a$
(\ref{Grc}).  The subscript $C$ in $G_{RC}$ emphasizes that the
resistance to crack growth has reached an asymptotic or
\textit{critical} value.  
%
% XXX this is new
%
This post \textit{R}-curve regime at constant resistance 
to crack growth obtained here in an analytical form [Eq.(\ref{Grc})] 
is in agreement with the one generally assumed in the phenomenological 
approaches (for instance \cite{Baz90}).  
%  
%The fracture behavior defined by
%Eq.(\ref{Gr}) and Eq.(\ref{Grc}) is in agreement with the
%phenomenological \textit{R}-curve proposed by Bazant and collaborators
%\cite{Baz90} in which a post \textit{R}-curve regime at constant
%resistance to crack growth is suggested.
%
% XXX this is new
%  
In Fig.~1, Eq.(\ref{Gr}) and Eq.(\ref{Grc}) are used to fit the data. 
Eq.(\ref{Gr}) provides a good description of the growth of the energy 
release rate. This fit needs to use an overestimated ratio 
$A/{l_o}^{1-{\zeta}_{loc}}$.    
Note that, a previous approach obtained from the energy released 
by a fractal pattern of microcracks ahead of the main crack of a 
quasi-brittle material \cite{Bor97} have shown equally a 
\textit{R}-curve behavior.  On the other hand, it has been shown that 
the roughness of a crack induces a decrease of the stress intensity 
factor \cite{Van97} and this shielding process is also known to be a 
source of the phenomenological \textit{R}-curves.  
These latter points emphasize that the connection between 
the anomalous roughening of fracture surfaces and fracture mechanics 
seems to reflect the particular fracture behavior of  
brittle materials showing toughening mechanisms.  

The main consequence of the anomalous scaling is the dependence of the
critical resistance $G_{RC}$ to crack growth on the specimen size $L$
[see Eq.(\ref{Grc})].  Size effect phenomena is one of the main
characteristic of the fracture behavior of heterogeneous brittle
materials such as concrete, rocks or wood \cite{Baz90}.
Figure~\ref{fig3} shows the theoretical evolution of critical 
resistance to crack growth as a function of the system size (or 
characteristic size) $L$ obtained for arbitrary values.  It appears 
that the size effect shows two asymptotic behaviors respectively 
$G_{RC}=2\gamma$ and $G_{RC}\sim L^{\zeta-{\zeta}_{loc}}$.  
From Eq.(\ref{Grc}), one can define the crossover length
$L_c=({l_o}^{1-{\zeta}_{loc}}/A)^{1/(\zeta-{\zeta}_{loc})}$. 
%
% XXX this is new
%
For small system sizes $L \ll L_c$, which correspond theoretically to 
"shallow" surfaces, \ie~surfaces that have a mean local angle of the 
crack profile smaller than $45$ degrees, there is no size effect :
$G_{RC}\simeq 2\gamma$.  The roughness of crack surfaces being 
negligible, real crack surfaces are weakly different
from the projected one and so, classical results of purely elastic 
fracture mechanics are recovered.  
For large system sizes $L \gg L_c$, the critical resistance evolves as 
a power law : $G_{RC}\sim L^{\zeta-{\zeta}_{loc}}$.  These fracture 
surfaces are theoretically "spiky" (\ie~the average local angle of 
the crack profile is greater than $45$ degrees) and in this case, 
the critical energy release rates are greater than the specific surface 
energy $2\gamma$. 
Note that the latter power law behavior is in good agreement with 
the observed size effect in wood as shown in Fig.~2.  In fact, the 
typical local angle of wood crack surfaces is smaller than $45$ 
degrees but the crossover length $L_c$, calculated with the ratio 
$A/{l_o}^{1-{\zeta}_{loc}}$ obtained from the \textit{R}-curve fit 
(Fig.~1), is smaller than the sizes $L$ of tested specimens.  

If the development of the roughness crack is driven by a Family-Vicsek
scaling instead of an anomalous scaling, the crack profile length
$\psi$ is independent of the crack increment $\Delta a$: $\psi (y)
\simeq \sqrt{ 1+ (A \: / \: {l_o}^{1-{\zeta}_{loc}})^2}\:L$.  Hence, 
the resistance to fracture growth reduces to :
\begin{equation}
        G_C (\Delta a) \simeq \: 2 \: \gamma \: 
        \sqrt{1+ \left( \frac{A}{{l_o}^{1-{\zeta}_{loc}}} \right)^2 } 
\label{GcFamily}     
\end{equation}
Thus, for a Family-Vicsek scaling, the resistance to fracture is
independent of the crack position $\Delta a$ and 
of the specimen size $L$.  This is consistent with the behavior of
purely elastic brittle materials. Likewise, if the global roughness
exponent tends to the local one ($\zeta \rightarrow {\zeta}_{loc}$),
the \textit{R}-curve behavior (\ref{Gr}) and the size effect (\ref{Grc}) 
vanish progressively and the fracture behavior is close to purely elastic 
brittle materials [Eq.(\ref{GcFamily})].
%
%%%%%%%%%%%%%%%%%%%%%%%%%%%%%%%%%%%%%%%%%%%%%%%%%%%%%%%%%%%%%%%%%%%%%%%%%%%%%%%%%%%%%%%%%
%

In conclusion, on the basis of complete descriptions (3D) of fracture 
surface morphologies given by Family-Vicsek and anomalous scaling 
laws, we have shown that the connections between the roughening of fracture 
surfaces and fracture mechanics are important. A Family-Vicsek scaling of 
the fracture roughness reflects a purely elastic brittle fracture behavior, 
while the fracture behavior obtained on the basis of an anomalous scaling 
accounts for a \textit{R}-curve behavior and size effect of the critical 
resistance to crack growth. Anomalous scaling reflects the experimental 
behavior of brittle heterogeneous materials showing toughening mechanisms.  
On going experimental studies attempt to link systematically mechanical 
behaviors and fracture roughening for different materials.
%
%
%%%%%%%%%%%%%%%%%%%%%%%%%%%%%%%%%%%%%%%%%%%%%%%%%%%%%%%%%%%%%%%%%%%%%%%%%%%%%%%%%%%%%%%%%
%

We are particularly greatful to J.R. Rice for fruitful discussions and 
enlighting remarks. This work was partly supported by the CNRS-ECODEV 
project.
%\newpage

\end{multicols}

%
%
%%%%%%%%%%%%%%%%%%%%%%%%%%%%%%%%%%%%%%%%%%%%%%%%%%%%%%%%%%%%%%%%%%%%%%%%%%%%%%%%%%%%%%%%%
%
%\newpage
%
\begin{figure}
\protect
\caption{\textit{R}-curve behavior observed for a TDCB fracture experiment 
in wood.  For this Norway Spruce specimen of characteristic size 
$L=30mm$, data are fitted
from Eq.(\ref{Gr}) and Eq.(\ref{Grc}) with 
the scaling exponent $(\zeta-{\zeta}_{loc})/z=0.32$ which is consistent 
with the expected result for this specimen :  
$(\zeta-{\zeta}_{loc})/z=0.36 \pm 0.09$.}
\label{fig1}
\end{figure}
\begin{figure}
\protect
\caption{Size effect on the critical energy release rate
$G_{RC}$ (Norway Spruce specimens, $11.3 \leq L \leq 60 mm$).  
The behavior
derived from the anomalous scaling (solid line): 
$G_{RC}\sim
L^{\zeta-{\zeta}_{loc}}$ [Eq.(\ref{Grc})] is characterized 
by the exponent $0.69$ which is in good agreement with the expected 
result for Spruce : $\zeta-{\zeta}_{loc}=0.73 \pm 0.17$.}
\label{fig2}
\end{figure}
\begin{figure}
\protect
\caption{Theoretical size effect on the critical resistance to crack 
growth [Eq.\ref{Grc}] obtained 
from an anomalous scaling and for the 
arbitrary scaling exponents : $\zeta=1.3$, ${\zeta}_{loc}=0.8$, $A=0.1$ 
and $l_o=1$.}
\label{fig3}
\end{figure}
%
%
%%%%%%%%%%%%%%%%%%%%%%%%%%%%%%%%%%%%%%%%%%%%%%%%%%%%%%%%%%%%%%%%%%%%%%%%%%%%%
%
%\psfig{figure=fig1.fig}
%
%
%\psfig{figure=fig2.eps}
%
\newpage
\begin{figure}
\psfig{figure=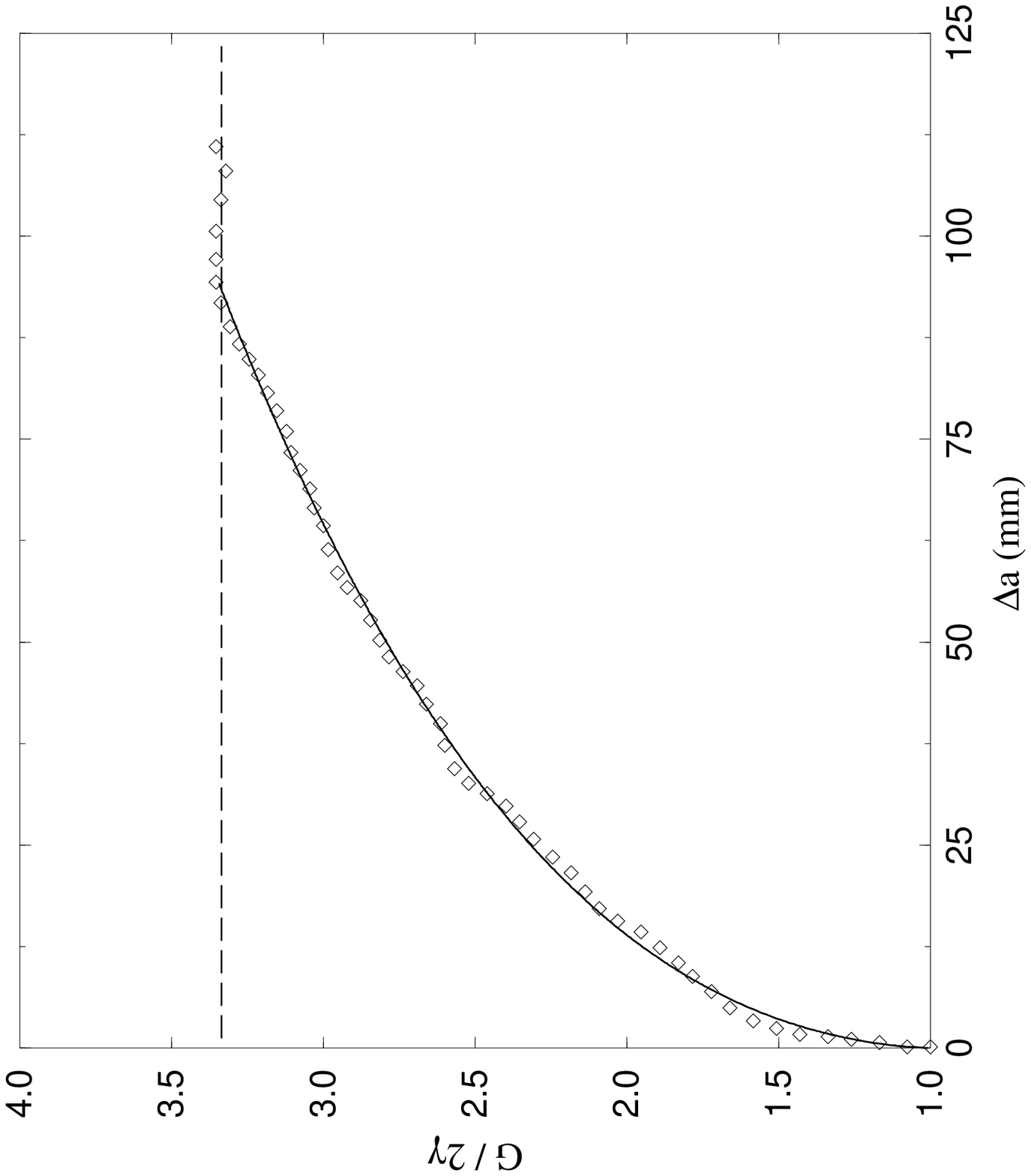,width=15cm,angle=0}
%\vspace*{1cm}
%\protect
%\caption{}
%\label{fig1}
\end{figure}
\clearpage
\begin{figure}
\psfig{figure=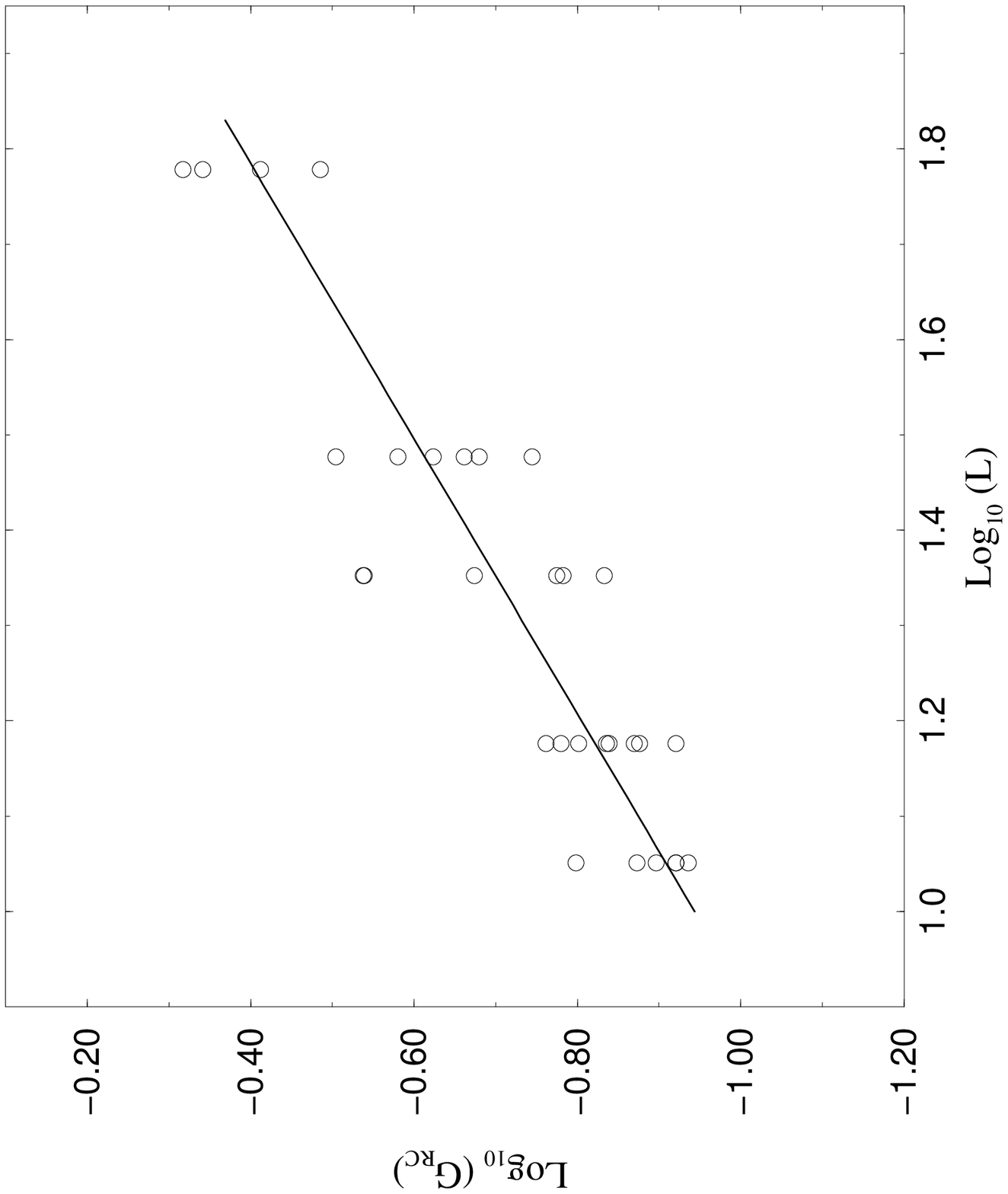,width=15cm,angle=0}
%\vspace*{1cm}
%\protect
%\caption{}
%\label{fig2}
\end{figure}
\clearpage
\begin{figure}
\psfig{figure=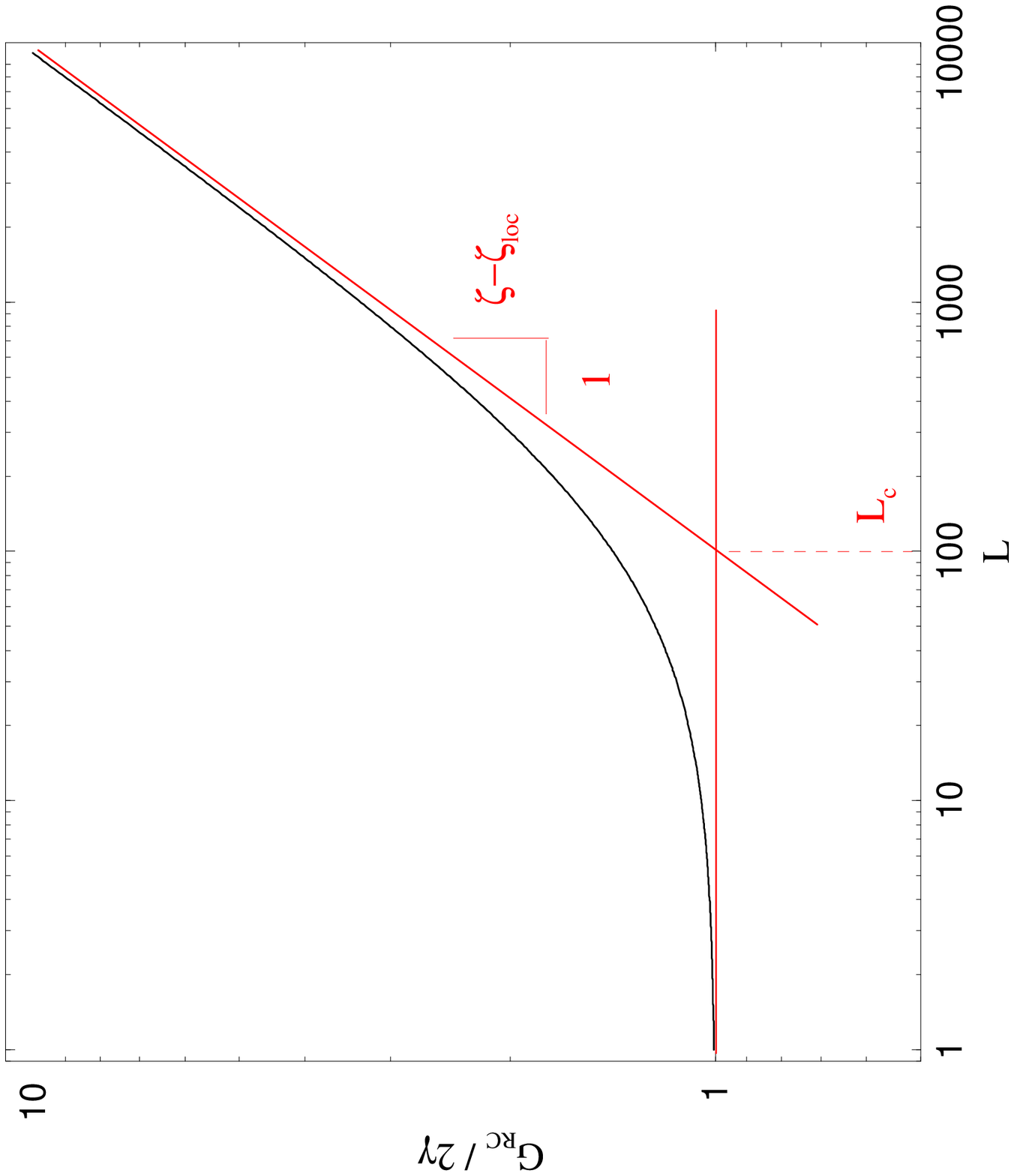,width=15cm,angle=0}
%\vspace*{1cm}
%\protect
%\caption{}
%\label{fig3}
\end{figure}
\end{document}